\documentstyle[aps,epsf,rotate,multicol]{revtex}

\begin{document}

\draft

\title{Economic Fluctuations and Diffusion} 

\author{Vasiliki Plerou$^{1,2}$, Parameswaran Gopikrishnan$^1$,  
Lu\'{\i}s~A.~Nunes Amaral$^1$,\\ 
Xavier Gabaix$^3$ and H. Eugene Stanley$^1$\\}

\address{$^1$Center for Polymer Studies and Department of Physics,\\
Boston University Boston, MA 02215 USA\\
$^2$Department of Physics, Boston College, Chestnut Hill, MA 02164 USA\\
$^3$Department of Economics, Massachusetts Institute of Technology, 
Cambridge, MA 02142 USA\\}

\date{\today}
\maketitle

\begin{abstract}

Stock price changes occur through transactions, just as diffusion in
physical systems occurs through molecular collisions.  We
systematically explore this analogy~\cite{Bachelier00} and quantify
the relation between trading activity --- measured by the number of
transactions $N_{\Delta t}$ --- and the price change $G_{\Delta t}$
for a given stock, over a time interval $[t,\,t+\Delta t]$.  To this
end, we analyze a database documenting every transaction for 1000 US
stocks over the two-year period 1994--1995~\cite{TAQ}. We find that
price movements are equivalent to a complex variant of diffusion,
where the diffusion coefficient fluctuates drastically in time. We
relate the analog of the diffusion coefficient to two microscopic
quantities: (i) the number of transactions $N_{\Delta t}$ in $\Delta
t$, which is the analog of the number of collisions and (ii) the local
variance $w^2_{\Delta t}$ of the price changes for all transactions in
$\Delta t$, which is the analog of the local mean square displacement
between collisions. We study the distributions of both $N_{\Delta t}$
and $w_{\Delta t}$, and find that they display power-law
tails. Further, we find that $N_{\Delta t}$ displays long-range
power-law correlations in time, whereas $w_{\Delta t}$ does not. Our
results are consistent with the interpretation that the pronounced
tails of the distribution of $G_{\Delta
t}$~\cite{Bouchaud,Farmer,Mandelbrot63,Mantegna95,Ghasghaie96,Pagan96,Lux96,Loretan94,Muller90,Gopi99}
are due to $w_{\Delta t}$, and that the long-range correlations
previously found~\cite{Ding93,Yanhui97,Lundin99,Andersen,Sornette} for
$\vert G_{\Delta t} \vert$ are due to $N_{\Delta t}$.

\end{abstract}

\begin{multicols}{2}

\bigskip
%\noindent
%\pacs{PACS numbers: 89.90.+n}

\begin{figure}
\narrowtext
\centerline{
\epsfysize=0.7\columnwidth{\rotate[r]{\epsfbox{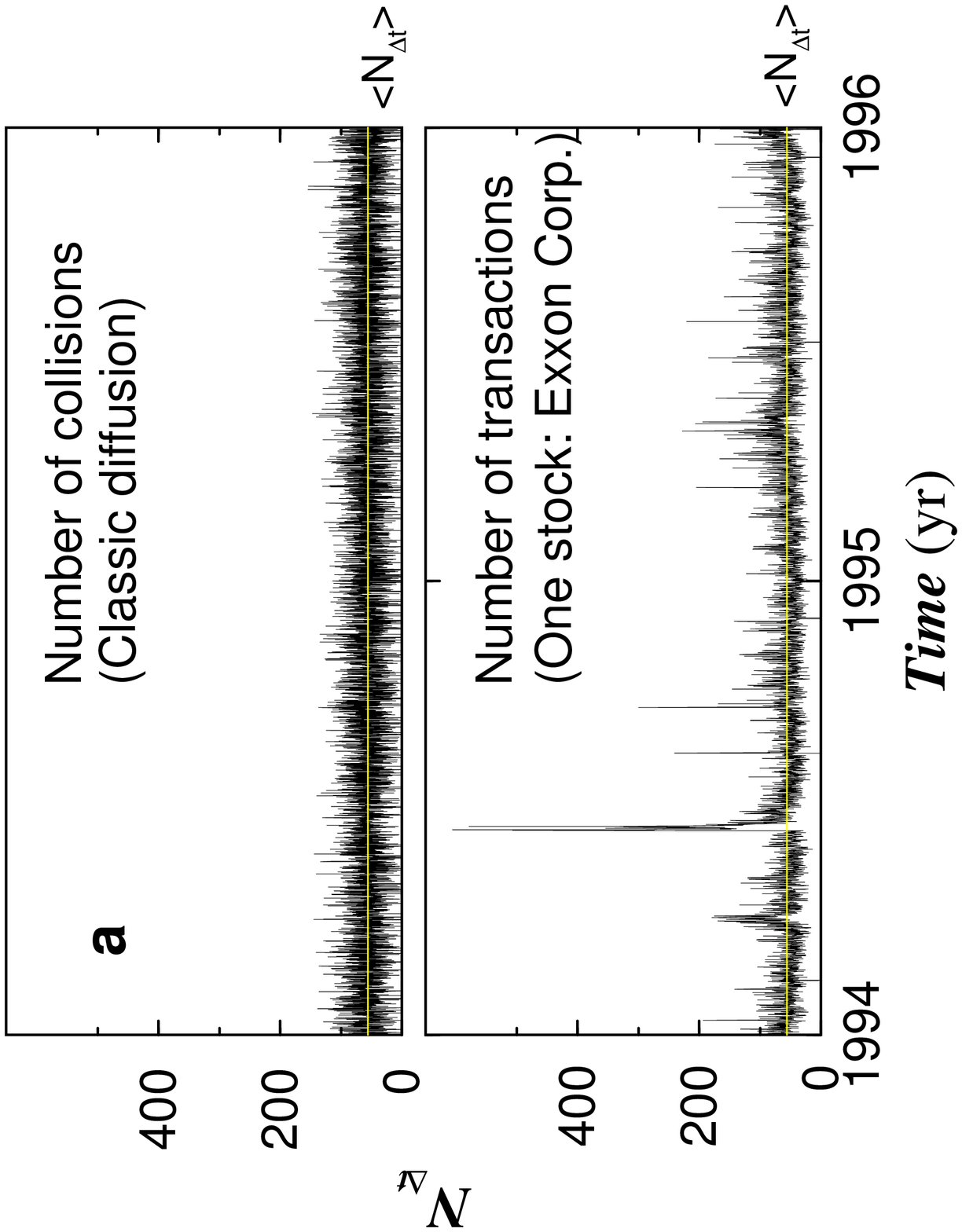}}}
}
\vspace{0.5cm}
\centerline{
\epsfysize=0.7\columnwidth{\rotate[r]{\epsfbox{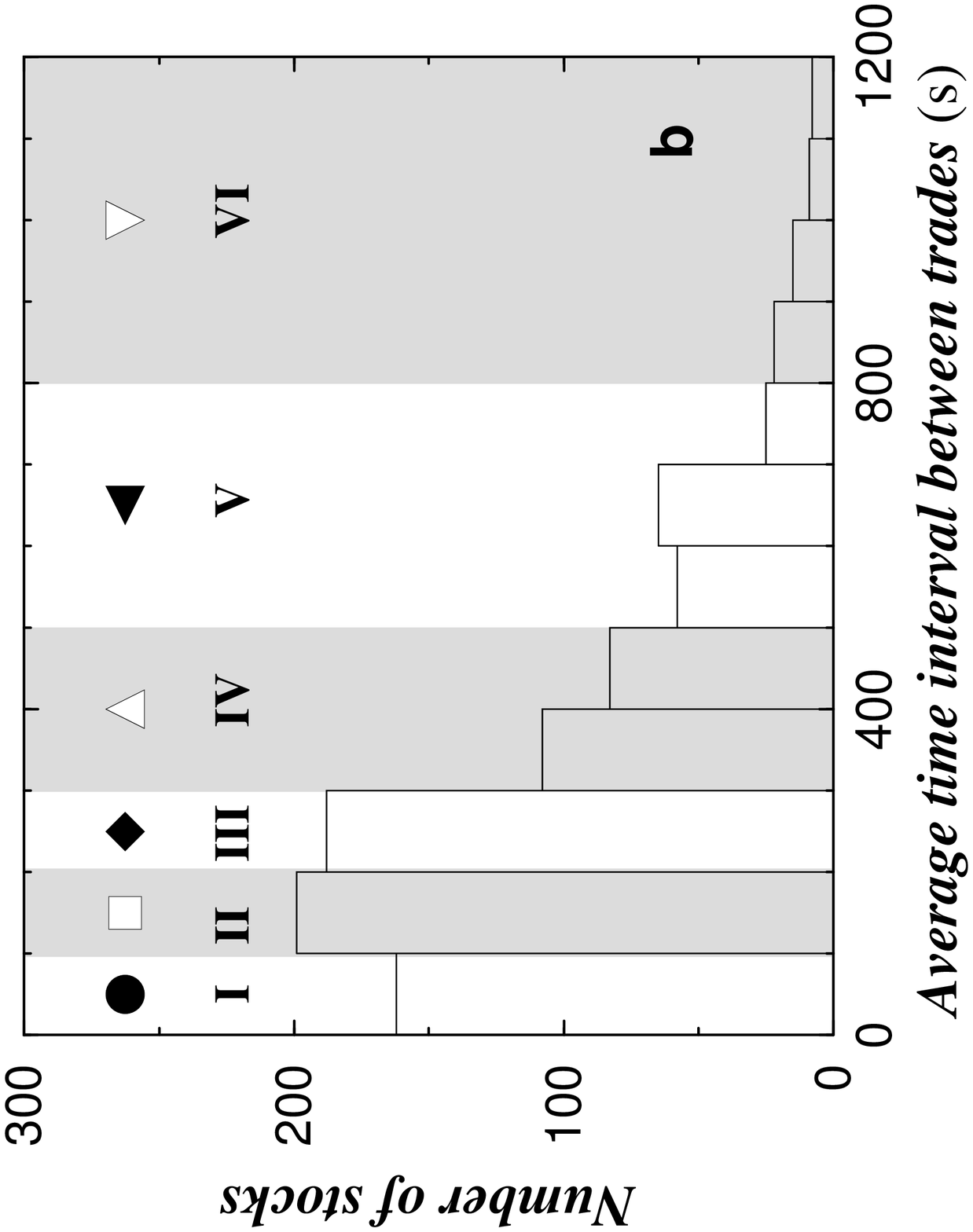}}}
}
\vspace{0.5cm}
\centerline{
\epsfysize=0.7\columnwidth{\rotate[r]{\epsfbox{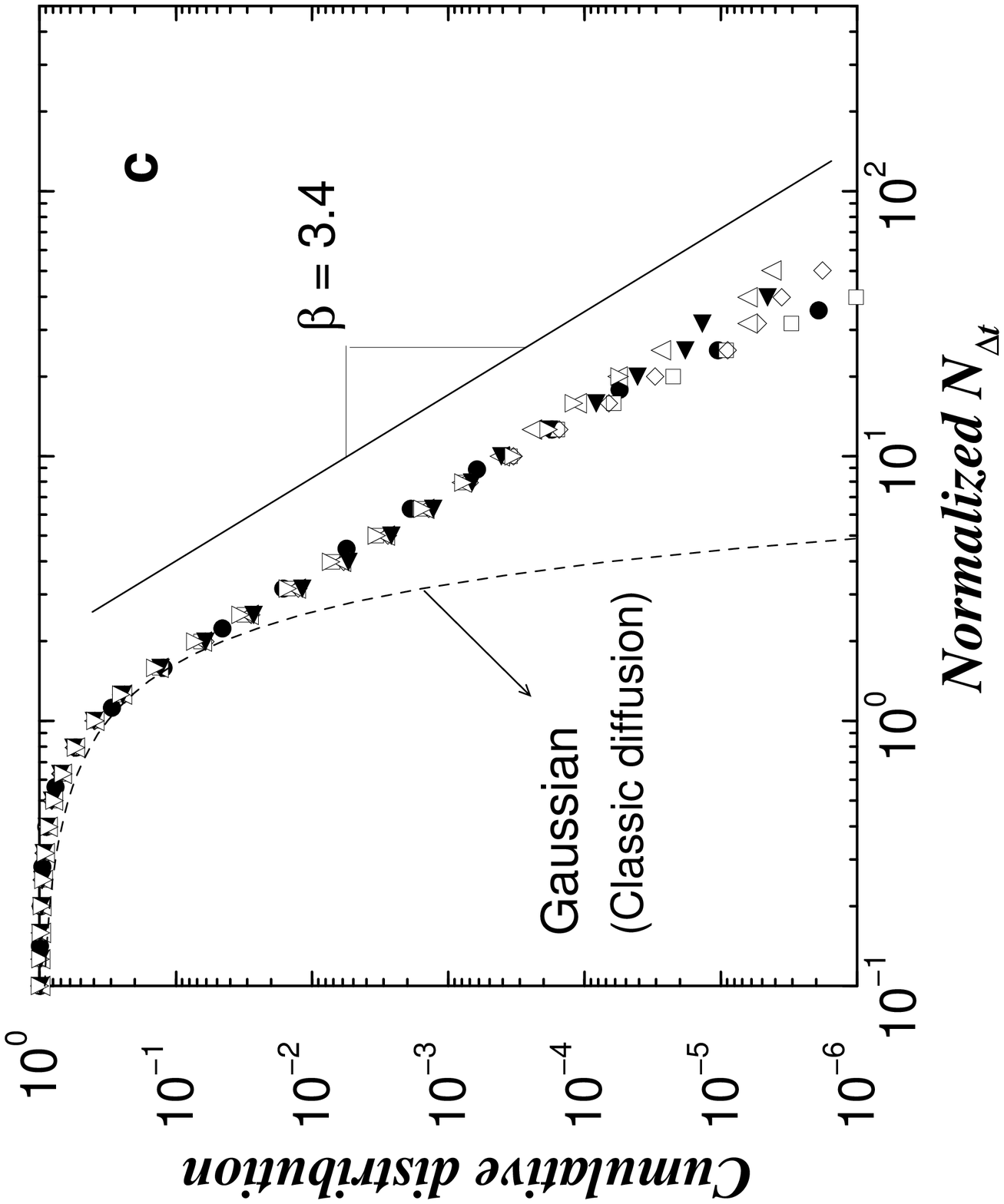}}}
}
\vspace{0.5cm}
\centerline{
\epsfysize=0.7\columnwidth{\rotate[r]{\epsfbox{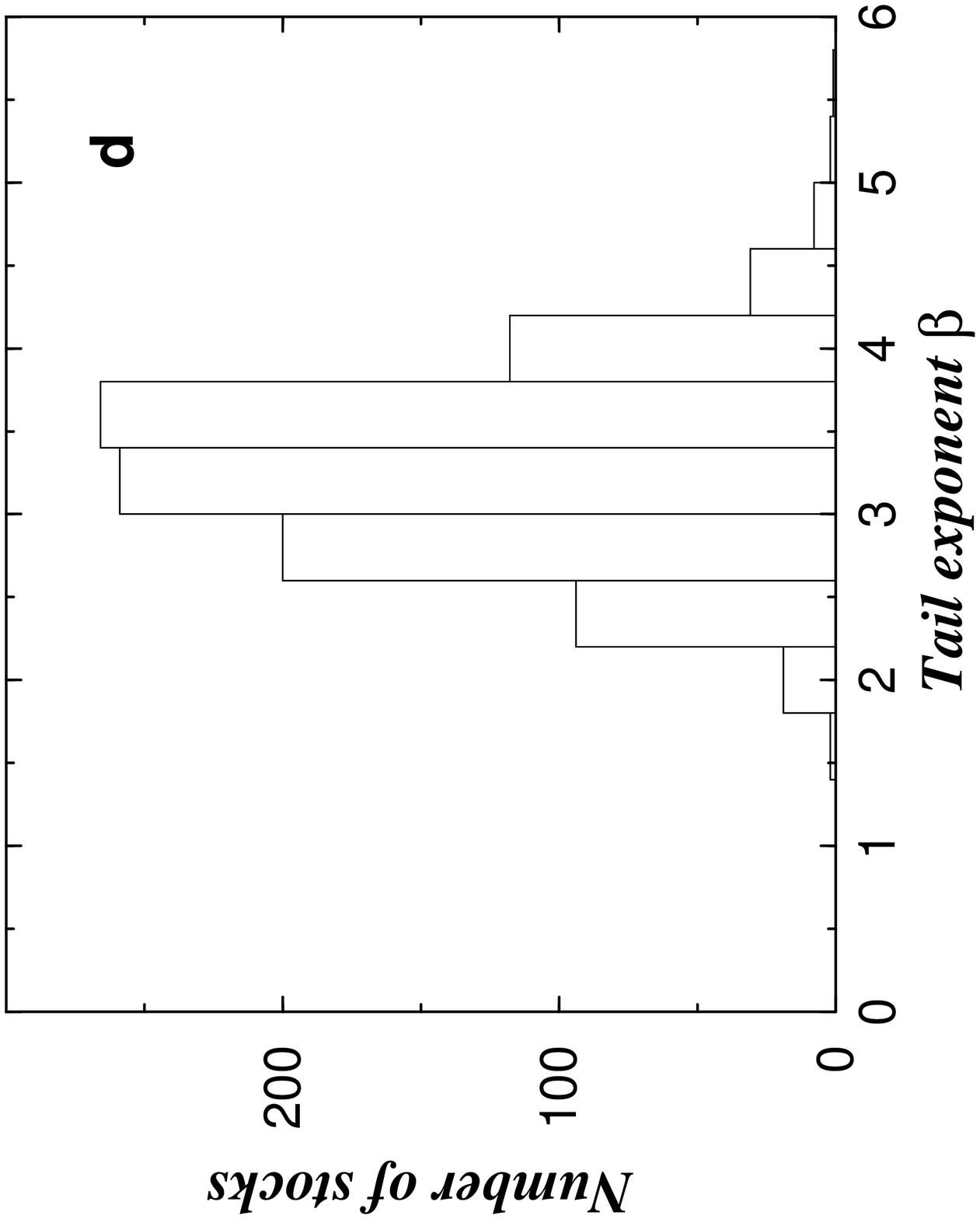}}}
}
\vspace{0.5cm}
\centerline{
\epsfysize=0.7\columnwidth{\rotate[r]{\epsfbox{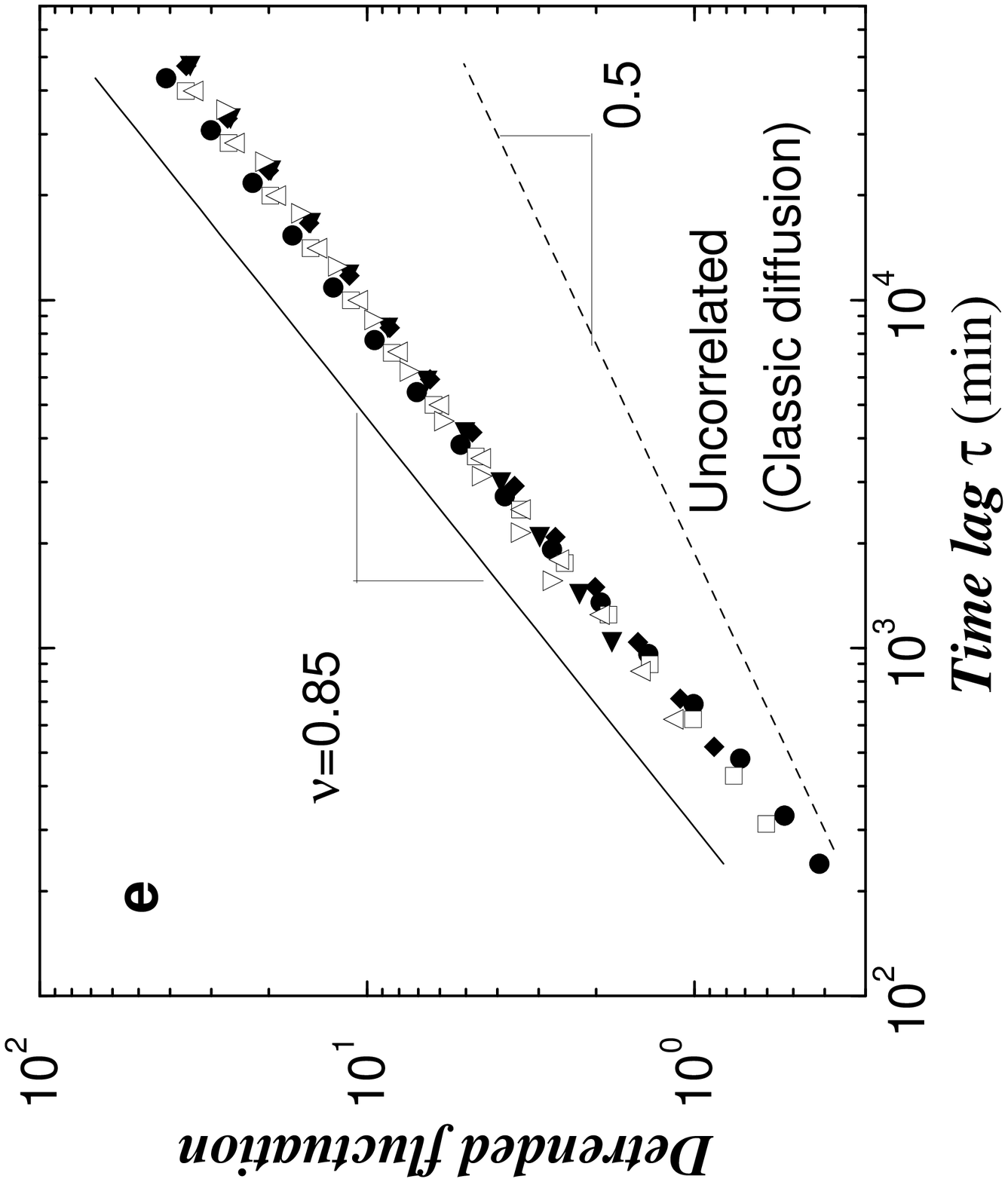}}}
}
\vspace{0.5cm}
\centerline{
\epsfysize=0.7\columnwidth{\rotate[r]{\epsfbox{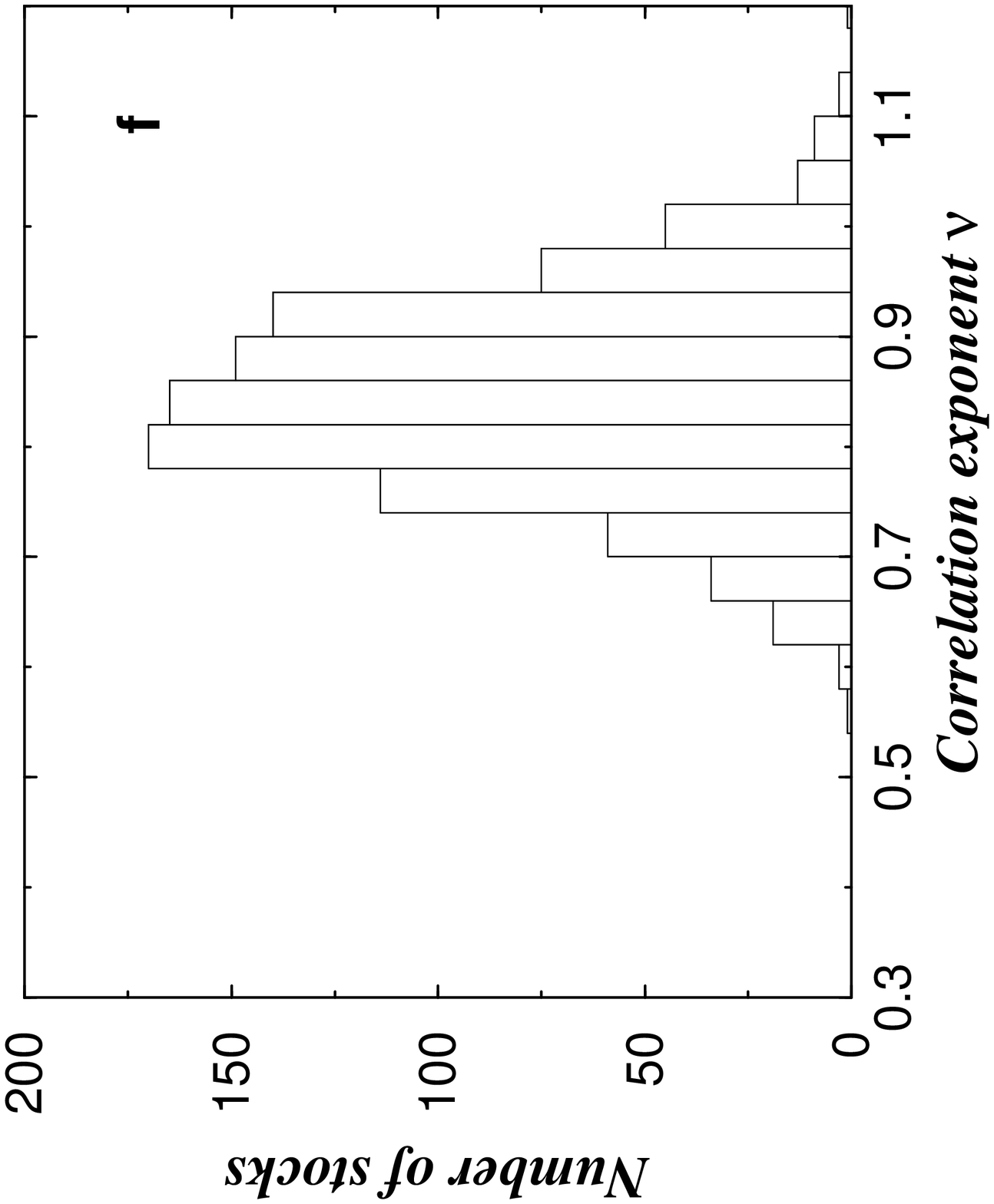}}}
}
\caption{ Statistical properties of $N_{\Delta t}$. {\bf a)} 
The lower panel shows $N_{\Delta t}$ for Exxon Corporation with
$\Delta t =30$~min and the average value $\langle N_{\Delta t}
\rangle \approx 52$. The upper panel shows a sequence of uncorrelated
Gaussian random numbers with the same mean and variance, which depicts
the number of collisions in $N_{\Delta t}$ for the classic diffusion
problem. Note that in contrast to diffusion, $N_{\Delta t}$ for Exxon
shows frequent large events of the magnitude of tens of standard
deviations, which would be forbidden for Gaussian statistics. {\bf b)}
The histogram of the average time interval between trades $\langle
\delta t \rangle$ for the 1000 stocks studied. In order to ensure that
the sampling time interval $\Delta t$ for each stock contains
sufficient number of transactions, we partition the stocks into 6
groups (I--VI) based on $\langle \delta t
\rangle$. For a specific group, we choose a sampling time $\Delta t$
at least 10 times larger than the average value of $\langle \delta t
\rangle$ for that group. We choose the sampling time interval 
$\Delta t = 30, 39, 65, 78, 130$ and $190$~min respectively for groups
I--VI. {\bf c)} Log-log plot of the cumulative distribution of
$N_{\Delta t}$ for the stocks in each of the six groups in {\bf
b)}. Since each stock has a different average value of $\langle
N_{\Delta t} \rangle$, we use a normalized number of transactions
$n_{\Delta t} \equiv N_{\Delta t}/\langle N_{\Delta t}
\rangle$. Each symbol shows the cumulative distribution $P\{n_{\Delta
t} > x\}$ of the normalized number of transactions $n_{\Delta t}$ for
all stocks in each group. {\bf d)} The histogram of exponents obtained
by fits to the cumulative distributions $P\{N_{\Delta t} > x\}$ for
each of the 1000 stocks. For the 1000 stocks studied, we obtain an
average value $\beta = 3.40$. We calculate an error estimate $\pm
0.05$ by dividing the standard deviation of the estimates of $\beta$
by $\protect\sqrt{1000}$.}
\label{fig.N}
\end{figure}

Consider the diffusion~\cite{Chandrasekhar,Montroll,Bouchaud90} of an
ink particle in water. Starting out from a point, the ink particle
undergoes a random walk due to collisions with the water
molecules. The distance covered by the particle after a time $\Delta
t$ is
\begin{mathletters}
\begin{equation}
x_{\Delta t} = \sum_{i=1}^{N_{\Delta t}} \, \delta x_i\,,
\label{sumdx}
\end{equation}
where $\delta x_i$ are the distances that the particle moves in
between collisions, and $N_{\Delta t}$ denotes the number of
collisions during the interval $\Delta t$. The distribution
$P(x_{\Delta t})$ is Gaussian with a variance $\langle x_{\Delta t}^2
\rangle = N_{\Delta t} \, w_{\Delta t}^2$, where the local mean square 
displacement $w_{\Delta t}^2 \equiv \langle(\delta x_i)^2 \rangle$ is
the variance of the individual steps $\delta x_i$ in the interval
$[t,\, t+\Delta t]$.

For the classic diffusion problem considered above: (i) the
probability distribution $P(N_{\Delta t})$ is a ``narrow'' Gaussian,
i.e., has a standard deviation much smaller than the mean $\langle
N_{\Delta t} \rangle$, (ii) the time between collisions of an ink
particle are not strongly correlated, so $N_{\Delta t}$ at any future
time $t+\,\tau$ depends at most weakly on $N_{\Delta t}$ at time
$t$---i.e., the correlation function $\langle N_{\Delta t} (t)
N_{\Delta t} (t + \tau) \rangle$ has a short-range exponential decay,
(iii) the distribution $P(w_{\Delta t})$ is also a narrow Gaussian,
(iv) the correlation function $\langle w_{\Delta t} (t) w_{\Delta t}
(t + \tau) \rangle$ has a short-range exponential decay and (v) the
variable $\epsilon \equiv x_{\Delta t}/(w_{\Delta t} \sqrt{N_{\Delta
t}})$ is uncorrelated and Gaussian-distributed. These conditions
result in $x_{\Delta t}$ being Gaussian distributed and weakly
correlated.

An ink particle diffusing under more general conditions would result
in a quite different distribution of $x_{\Delta t}$, such as in a
bubbling hot spring, where the characteristics of bubbling depend on a
wide range of time and length scales. In the following, we will
present empirical evidence that the movement of stock prices is
equivalent to a complex variant of classic diffusion, specified by the
following conditions: (i) $P(N_{\Delta t})$ is not a Gaussian, but has
a power-law tail, (ii) $N_{\Delta t}$ has long-range power-law
correlations, (iii) $P(w_{\Delta t})$ is not a Gaussian, but has a
power-law tail, (iv) the correlation function $\langle w_{\Delta t}
(t) w_{\Delta t} (t + \tau) \rangle$ is short ranged, and (v) the
variable $\epsilon \equiv x_{\Delta t}/(w_{\Delta t}
\sqrt{N_{\Delta t}})$ is Gaussian distributed and short-range correlated.
Under these conditions, the statistical properties of $x_{\Delta t}$
will depend on the exponents characterizing these power laws.

Just as the displacement $x_{\Delta t}$ of a diffusing ink particle is
the sum of $N_{\Delta t}$ individual displacements $\delta x_i$, so
also the stock price change $G_{\Delta t}$ is the sum of the price
changes $\delta p_i$ of the $N_{\Delta t}$ transactions in the
interval $[t, t+\Delta t]$,
\begin{equation}
G_{\Delta t} = \sum_{i=1}^{N_{\Delta t}} \delta p_i\,.
\label{sumdp}
\end{equation}
Figure~1a shows $N_{\Delta t}$ for classic diffusion and for one stock
(Exxon Corporation). The number of trades for Exxon displays several
events the size of tens of standard deviations and hence is
inconsistent with a Gaussian
process~\cite{Clark73,Mandelbrot67,Epps76,Tauchen83,Stock88,Guillaume}.

(i) We first analyze the distribution of $N_{\Delta
t}$. Figure~\ref{fig.N}c shows that the cumulative distribution of
$N_{\Delta t}$ displays a power-law behavior $P\{N> x\}\sim
x^{-\beta}$. For the 1000 stocks analyzed, we obtain a mean value
$\beta = 3.40 \pm 0.05$ (Fig.~\ref{fig.N}d). Note that $\beta > 2$ is
outside the L\'evy stable domain $0 < \beta < 2$.
\end{mathletters}

(ii) We next determine the correlations in $N_{\Delta t}$. We find
that the correlation function $\langle N_{\Delta t}(t) N_{\Delta t}(t+
\tau) \rangle$ is not exponentially decaying as in the case of classic
diffusion, but rather displays a power-law decay
(Fig.~\ref{fig.N}e,f). This result quantifies the qualitative fact
that if the trading activity ($N_{\Delta t}$) is large at any time, it
is likely to remain so for a considerable time thereafter.

\begin{figure}
\narrowtext
\centerline{
\epsfysize=0.7\columnwidth{\rotate[r]{\epsfbox{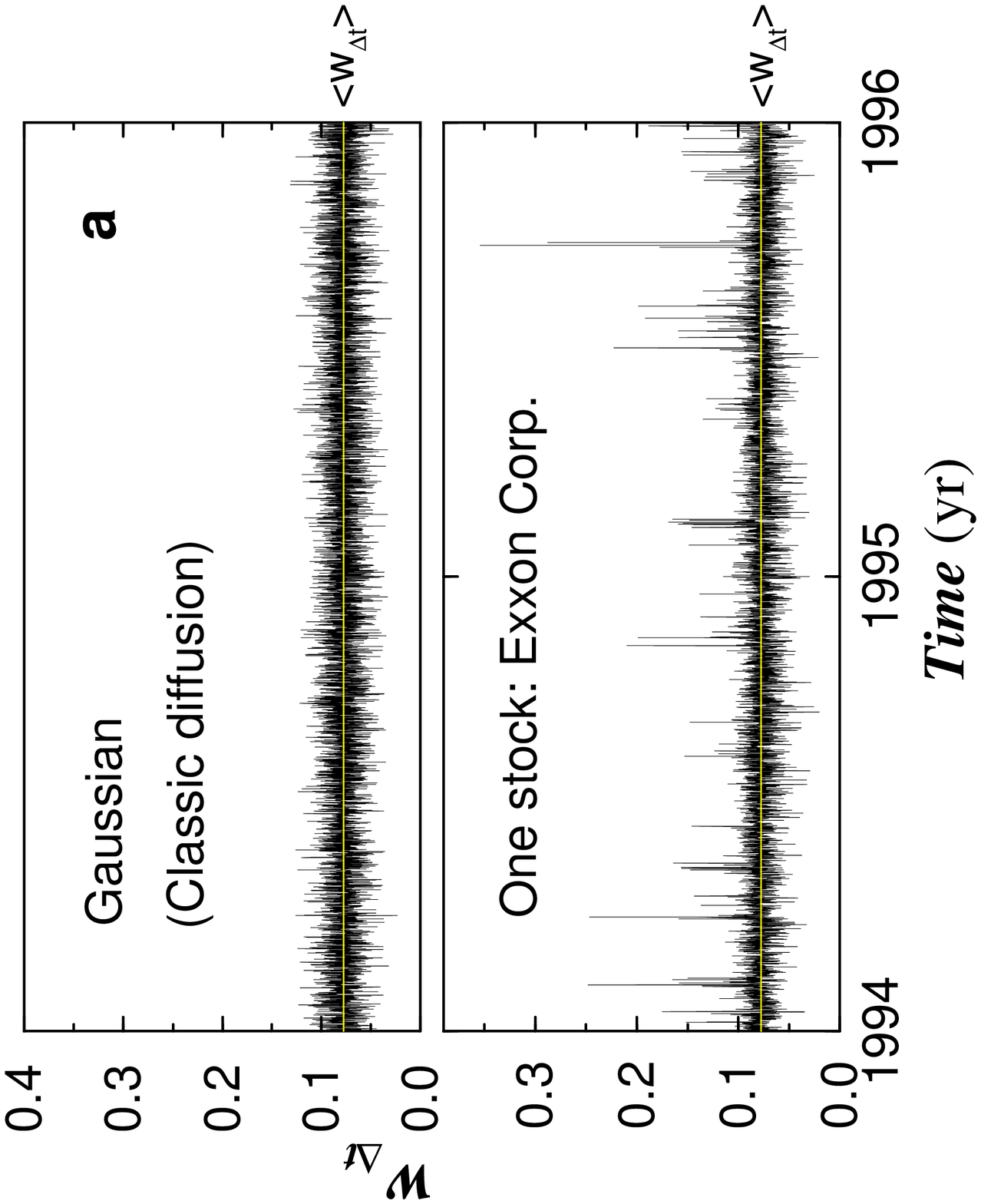}}}
}
\vspace{0.5cm}
\centerline{
\epsfysize=0.7\columnwidth{\rotate[r]{\epsfbox{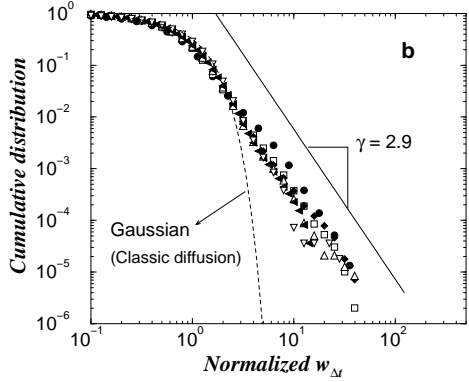}}}
}
\vspace{0.5cm}
\centerline{
\epsfysize=0.7\columnwidth{\rotate[r]{\epsfbox{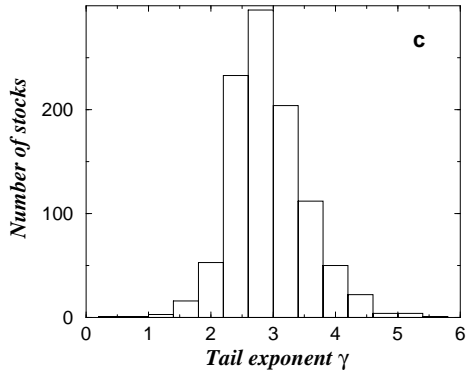}}}
}
\vspace{0.5cm}
\centerline{
\epsfysize=0.7\columnwidth{\rotate[r]{\epsfbox{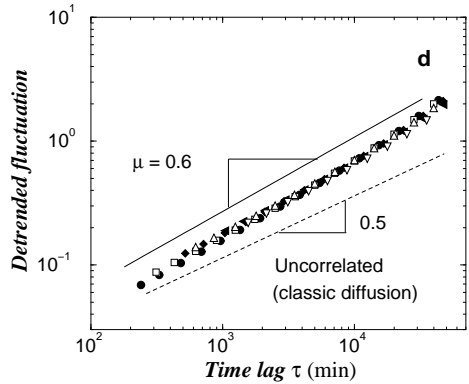}}}
}
\vspace{0.5cm}
\centerline{
\epsfysize=0.7\columnwidth{\rotate[r]{\epsfbox{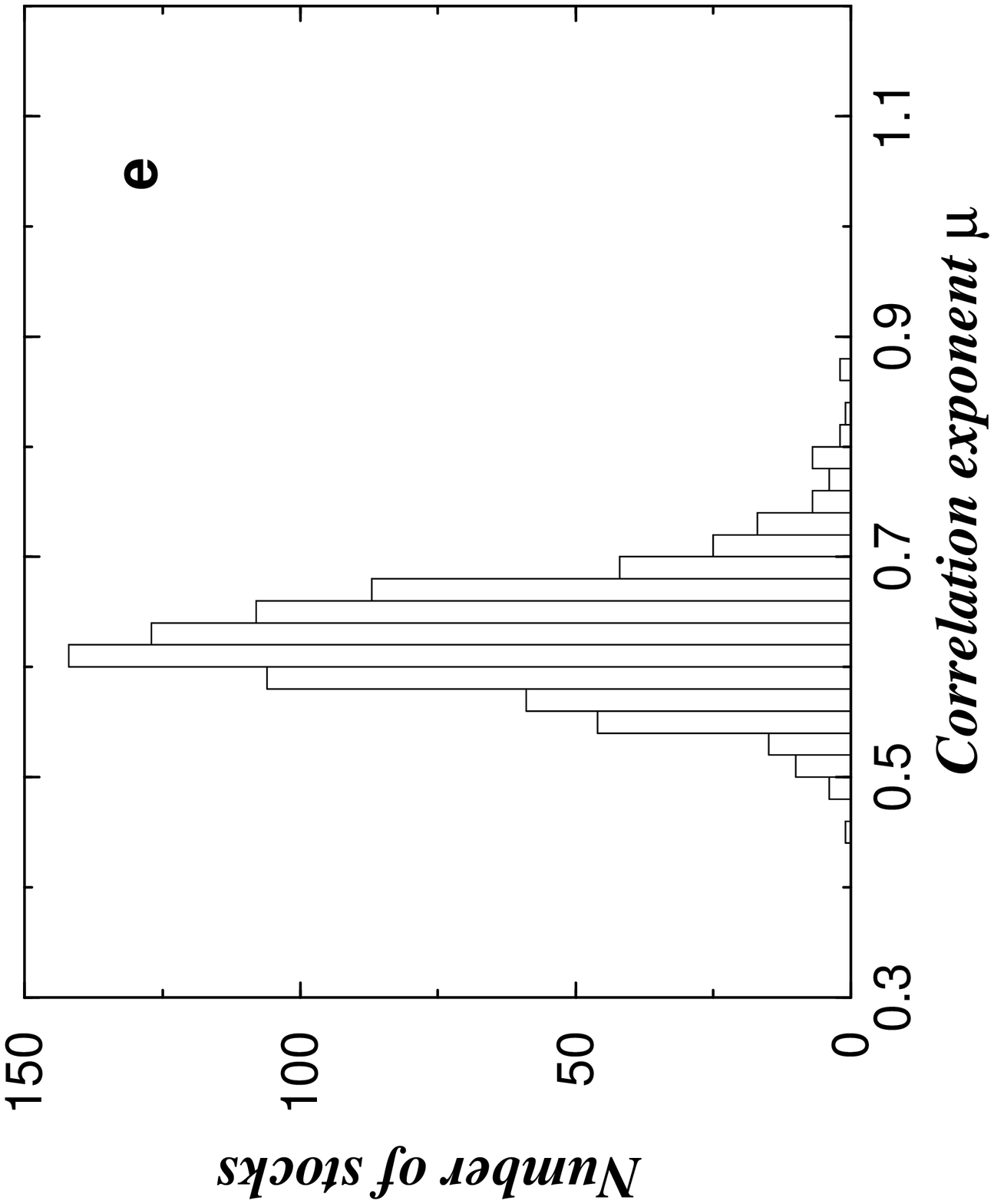}}}
}
\vspace{0.5cm}
\caption{Statistical properties of $w_{\Delta t}$. {\bf a)} The local 
standard deviation $w_{\Delta t}$ computed from price changes $\delta
p_i$ due to every transaction in the interval $[t,\,t+\Delta t]$ for
Exxon Corporation (lower panel) in contrast to uncorrelated Gaussian
random numbers with the same mean value $\langle w_{\Delta t} \rangle
\approx 0.08$ and variance (upper panel). The time series of $w_{\Delta t}$ for
Exxon shows a number of large events of the size of tens of standard
deviations. Intervals having fewer than 4 transactions are not used
for computing $w_{\Delta t}$. Note that the large values of $N_{\Delta
t}$ in Fig.~\protect\ref{fig.N}a do not correspond to large values of
$w_{\Delta t}$, showing that $N_{\Delta t}$ and $w_{\Delta t}$ are
weakly, if at all, correlated. {\bf b)} Log-log plot of the cumulative
distribution of $w_{\Delta t}$ for each of the six groups defined in
Fig.~\protect\ref{fig.N}b. Since the average value $\langle w_{\Delta
t} \rangle$ changes from one stock to another, we normalize $w_{\Delta
t}$ by $\langle w_{\Delta t} \rangle$. Each symbol shows the
cumulative distribution of the normalized $w_{\Delta t}$ for all
stocks in each group. {\bf c)} The power law exponents for the
cumulative distribution of $w_{\Delta t}$ obtained by fits to the
cumulative distributions of each of the 1000 stocks separately. We
obtain an average value $\gamma=2.9 \pm 0.1$.  {\bf d)} Log-log plot
of the detrended fluctuation $F(\tau)$ as a function of the time lag
$\tau$. Each symbol shows $F(\tau)$ averaged over all stocks in each
group. {\bf e)} The histogram of detrended fluctuation exponents
obtained by fitting $F(\tau)$ for each stock separately. We obtain an
average value $\mu= 0.60 \pm 0.01$.}
\label{fig.w}
\end{figure}

(iii) We then compute the variance $w_{\Delta t}^2\equiv\langle
(\delta p_i)^2 \rangle$ of the individual changes $\delta p_i$ due to
the $N_{\Delta t}$ transactions in the interval $[t,\,t+\Delta t]$
(Fig.~\ref{fig.w}a). We find that the distribution $P(w_{\Delta t})$
displays a power-law decay $P\{w_{\Delta t} > x\} \sim x^{-\gamma}$
(Fig.~\ref{fig.w}b). For the 1000 stocks analyzed, we obtain a mean
value of the exponent $\gamma = 2.9 \pm 0.1$ (Fig.~\ref{fig.w}c).

(iv) Next, we quantify correlations in $w_{\Delta t}$. We find that
the correlation function $\langle w_{\Delta t}(t)\, w_{\Delta t}
(t+\tau) \rangle$ shows only weak correlations
(Fig.~\ref{fig.w}d,e). This means that $w_{\Delta t}$ at any future
time $t+\tau$ depends at most weakly on $w_{\Delta t}$ at time $t$.

\begin{figure}
\narrowtext
\vspace{0.5cm}
\centerline{
\epsfysize=0.7\columnwidth{\rotate[r]{\epsfbox{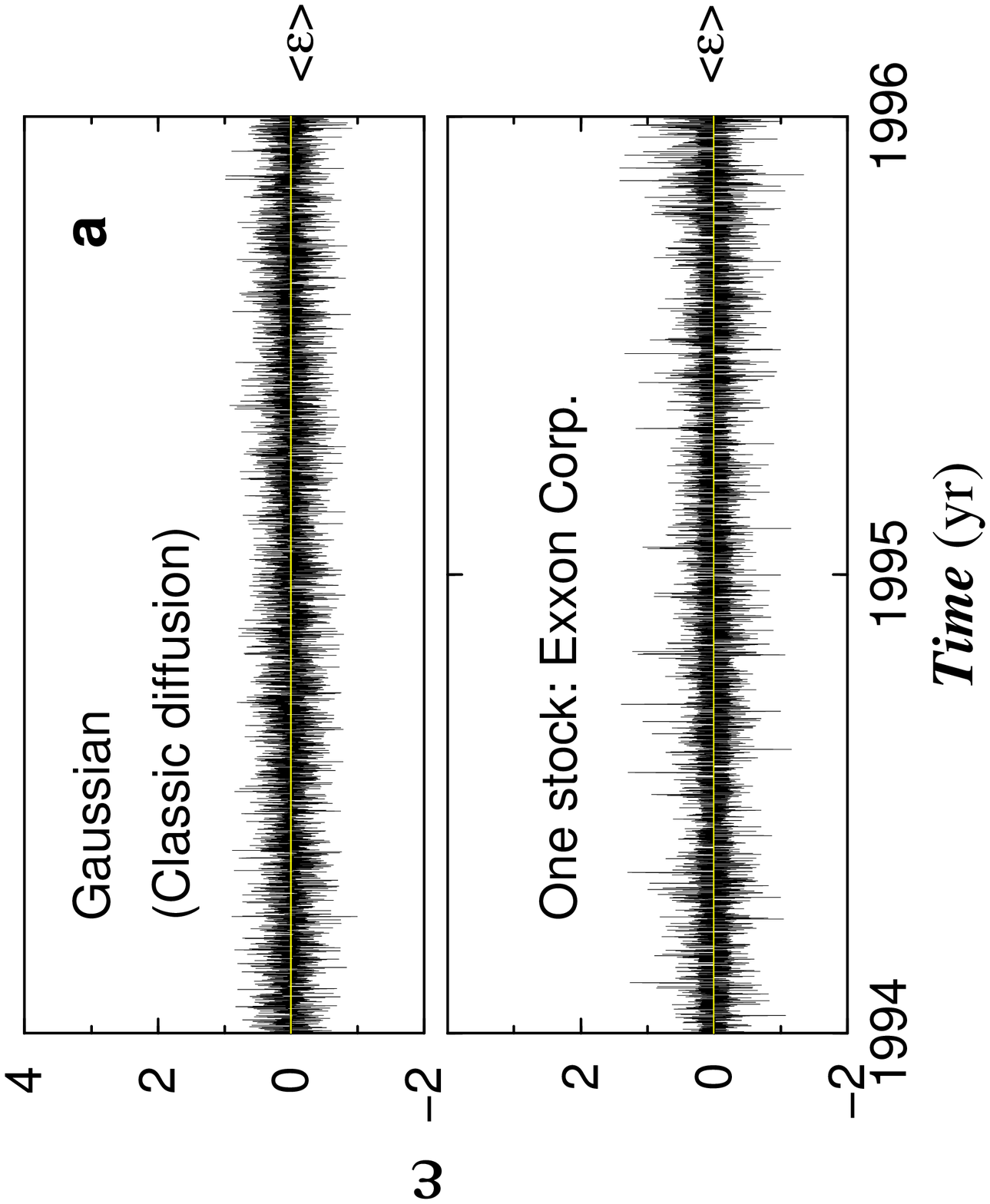}}}
}
\vspace{0.5cm}
\centerline{
\epsfysize=0.7\columnwidth{\rotate[r]{\epsfbox{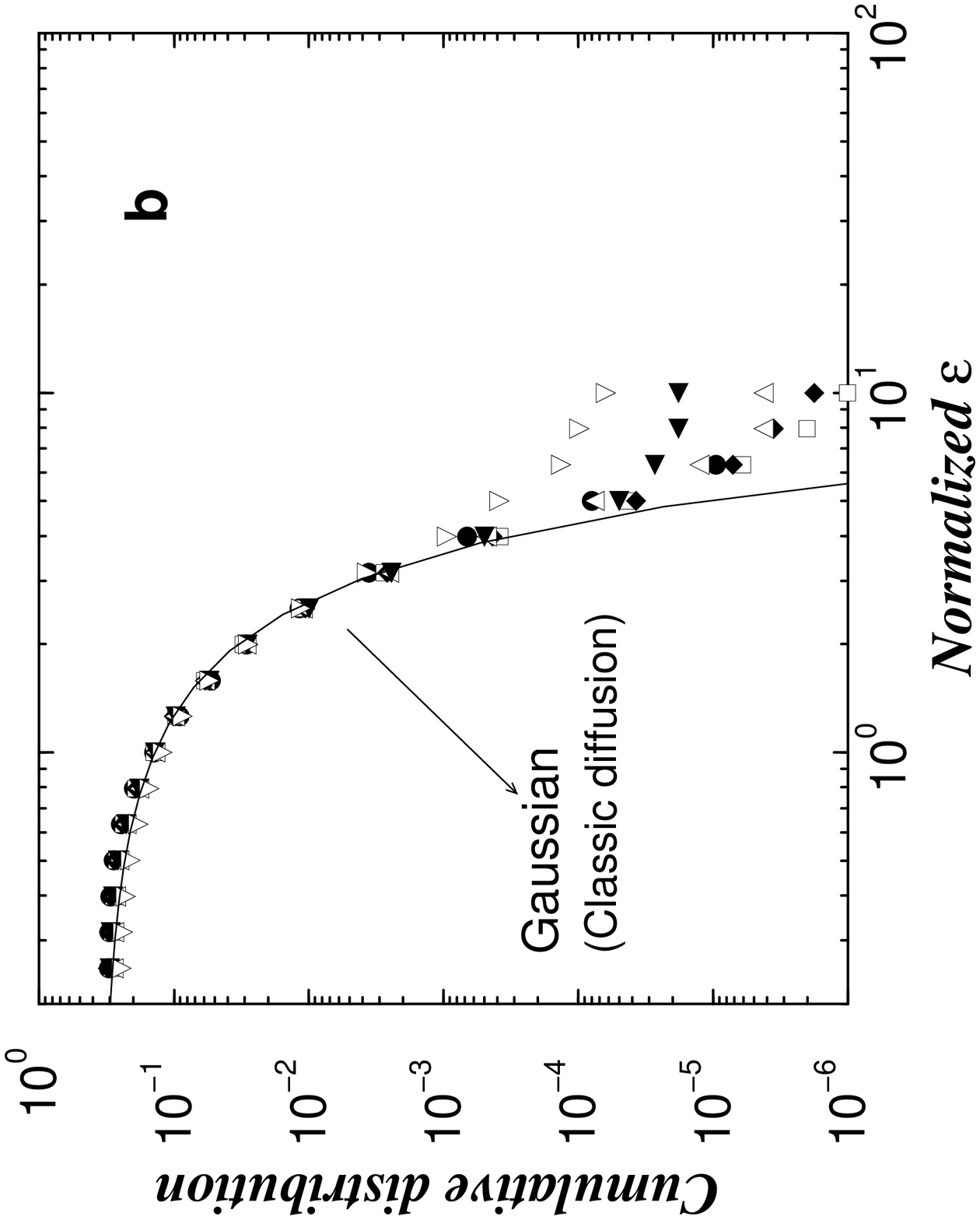}}}
}
\vspace{0.5cm}
\centerline{
\epsfysize=0.7\columnwidth{\rotate[r]{\epsfbox{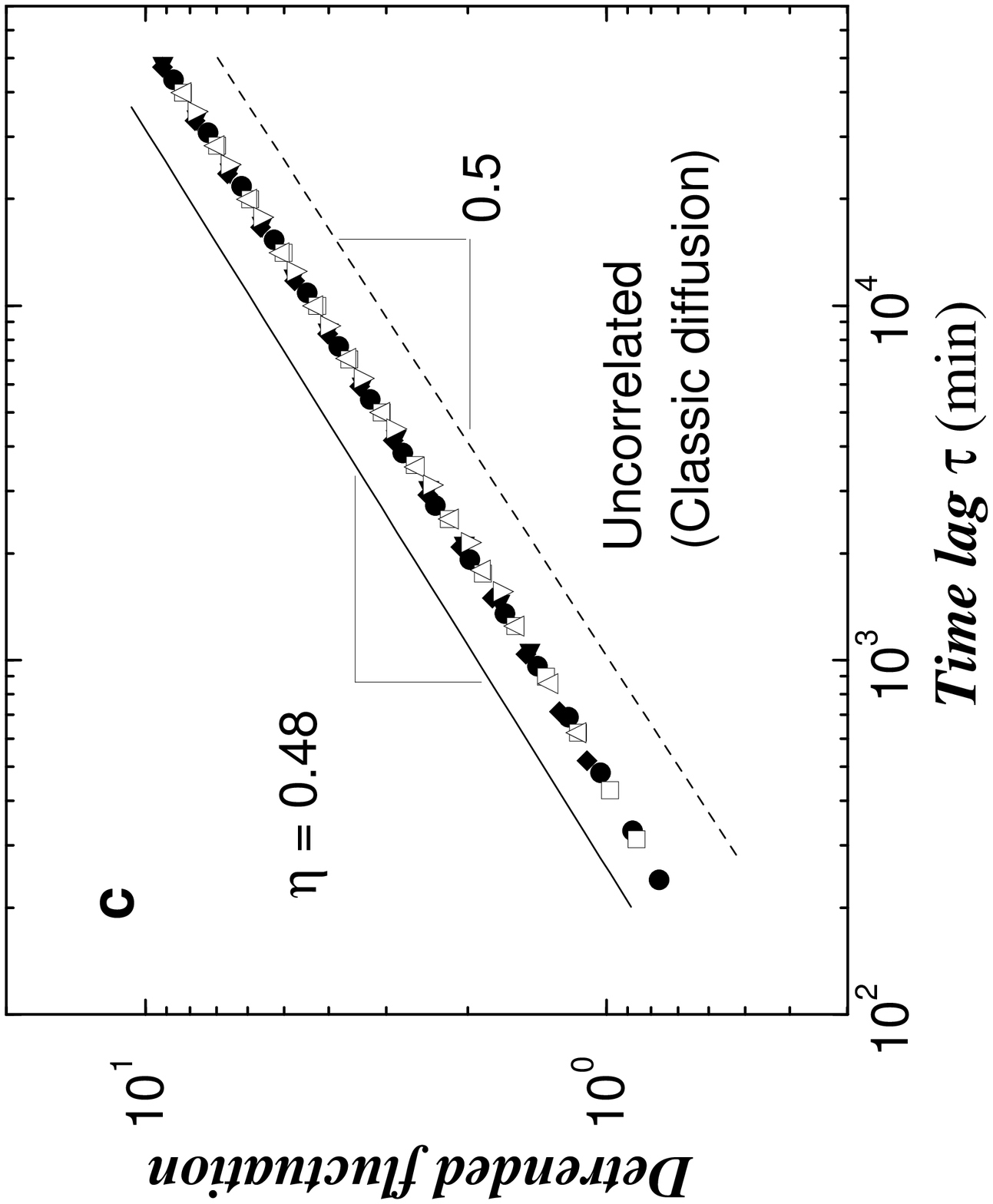}}}
}
\vspace{0.5cm}
\centerline{
\epsfysize=0.7\columnwidth{\rotate[r]{\epsfbox{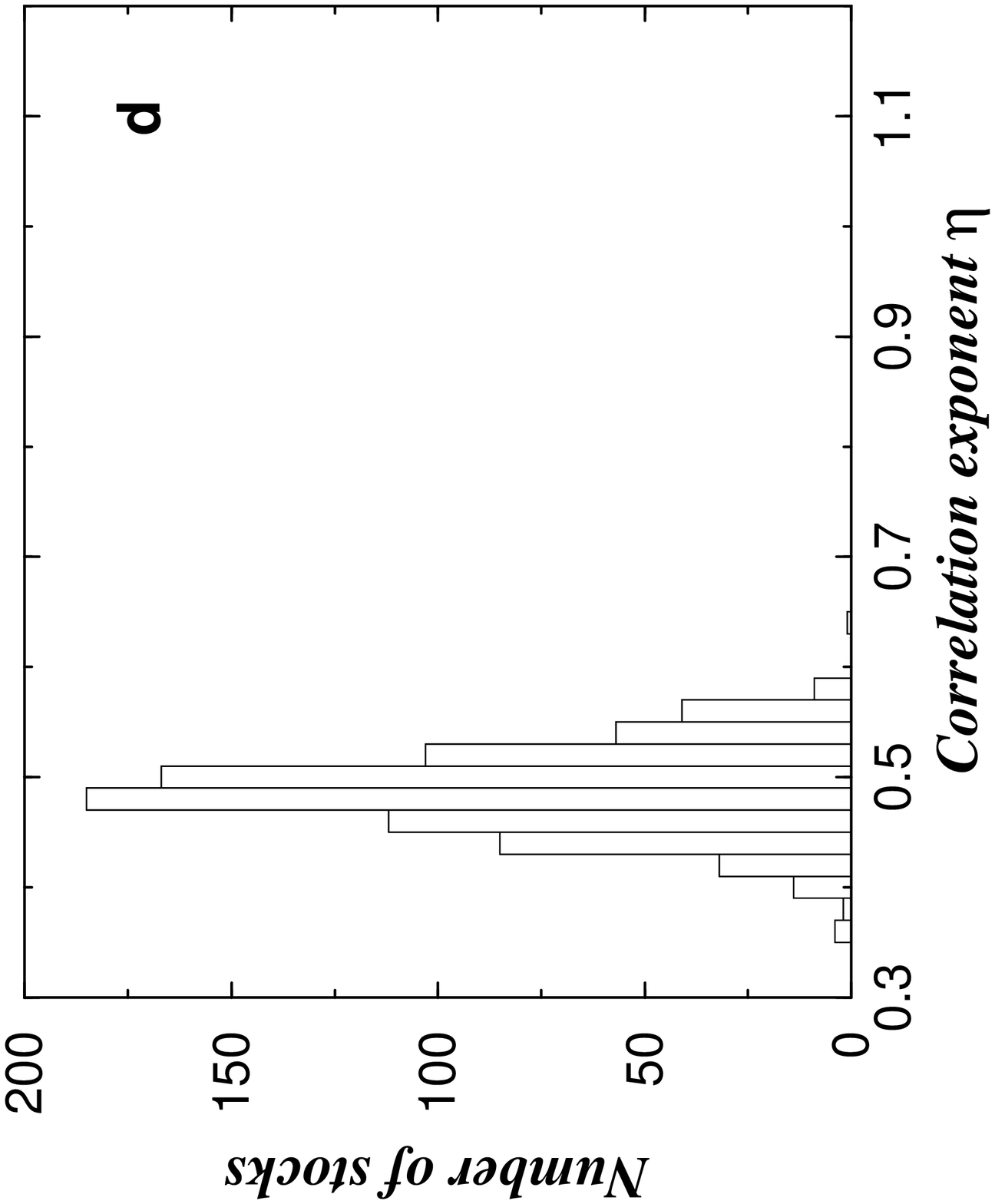}}}
}
\caption{Statistical properties of $\epsilon$. {\bf a)} The time series
of $\epsilon \equiv G_{\Delta t}/(w_{\Delta t} \protect\sqrt{N_{\Delta
t}})$ for Exxon Corporation (lower panel) in contrast with a sequence
of uncorrelated Gaussian random numbers with the same mean and
variance which depicts $x_{\Delta t}/(w_{\Delta t}
\protect\sqrt{N_{\Delta t}})$ for classic diffusion (upper
panel). {\bf b)} The positive tail of the cumulative distribution of
$\epsilon$ for the six groups. We normalize $\epsilon$ by its standard
deviation in order to compare different stocks. Each symbol shows the
cumulative distributions of the normalized $\epsilon$ for all stocks
in each of the six groups. The negative tail (not shown) displays
similar behavior. {\bf c)} Log-log plot of the detrended fluctuation
$F(\tau)$ averaged for all stocks belonging to each of the six groups.
{\bf d)} The histogram of detrended fluctuation exponents obtained by
fits to $F(\tau)$ for each stock. We obtain the average value
$\eta=0.48 \pm 0.06$.}
\label{fig.e}
\end{figure}

(v) Consider now $\delta p_i$ chosen only from the interval
$[t,\,t+\Delta t]$, and let us hypothesize that {\it these\/} $\delta
p_i$ are mutually independent and with a common distribution $P(\delta
p_i \vert t\in[t,t+\Delta t])$ having a finite variance $w_{\Delta
t}^2$. Under this hypothesis, the central limit theorem implies that
the ratio
\begin{equation}
\epsilon  \equiv {G_{\Delta t} \over w_{\Delta t} \sqrt{N_{\Delta t}}}
\label{eq.sub}
\end{equation}
must be a Gaussian-distributed random variable with zero mean and unit
variance~\cite{Feller}. Indeed, for classic diffusion, $x_{\Delta
t}/(w_{\Delta t}\sqrt{N_{\Delta}})$ is Gaussian-distributed and
uncorrelated (Fig.~\ref{fig.e}a). We confirm this hypothesis by
analyzing (a) the distribution $P(\epsilon)$, which we find to be
consistent with Gaussian behavior (Fig.~\ref{fig.e}b), and (b) the
correlation function $\langle \epsilon(t)\,\epsilon(t+\tau) \rangle$,
for which we find only short-range correlations (Fig.~\ref{fig.e}c,d).

Thus far, we have seen that the data for stock price movements support
the following results: (i) the distribution of $N_{\Delta t}$ decays
as a power-law, (ii) $N_{\Delta t}$ has long-range correlations, (iii)
the distribution of $w_{\Delta t}$ decays as a power-law, (iv)
$w_{\Delta t}$ displays only weak correlations, and (v) the price
change $G_{\Delta t}$ at any time is consistent with a
Gaussian-distributed random
variable~\cite{Clark73,Mandelbrot67,Epps76,Tauchen83,Engle95,Ghysels96,Jones94,Ane00}
with a time-dependent variance $N_{\Delta t} \, w_{\Delta t}^2$, that
is, the variable $\epsilon
\equiv G_{\Delta t}/(w_{\Delta t} \sqrt{N_{\Delta t}})$ is 
Gaussian-distributed and uncorrelated.

Next, we explore the implications of our empirical findings. Namely,
we show how the statistical
properties~\cite{Mandelbrot63,Mantegna95,Ghasghaie96,Pagan96,Lux96,Loretan94,Muller90,Gopi99,Ding93,Yanhui97,Lundin99}
of price changes $G_{\Delta t}$ can be understood in terms of the
properties of $N_{\Delta t}$ and $w_{\Delta t}$. We will argue that
the pronounced tails of the distribution of price changes
\cite{Mandelbrot63,Mantegna95,Ghasghaie96,Pagan96,Lux96,Loretan94,Muller90,Gopi99}
are largely due to $w_{\Delta t}$ and the long-range correlations
previously found~\cite{Ding93,Yanhui97,Lundin99,Andersen,Sornette} for $\vert
G_{\Delta t} \vert$ are largely due to the long-range correlations in
$N_{\Delta t}$. By contrast, in classic diffusion $N_{\Delta t}$ and
$w_{\Delta t}$ do not change the Gaussian behavior of $x_{\Delta t}$
because they have only uncorrelated Gaussian-fluctuations
\cite{Clark73,Feller}.

Consider first the distribution of price changes $G_{\Delta t}$, which
decays as a power-law $P\{G_{\Delta t} >x\} \sim x^{-\alpha}$ with an
exponent $\alpha \approx
3$~\cite{Pagan96,Lux96,Loretan94,Muller90,Gopi99}. Above, we reported
that the distribution $P\{N_{\Delta t}>x\} \sim x^{-\beta}$ with
$\beta\approx 3.4$ (Fig.~\ref{fig.N}c,d). Therefore,
$P\{\sqrt{N_{\Delta t}} > x\} \sim x^{-2\beta}$ with $2 \beta \approx
6.8$. Equation~(\ref{eq.sub}) then implies that $N_{\Delta t}$ alone
cannot explain the value $\alpha \approx 3$. Instead, $\alpha \approx
3$ must arise from the distribution of $w_{\Delta t}$, which indeed
decays with approximately the same exponent
$\gamma\approx\alpha\approx 3$ (Fig.~\ref{fig.w}b,c). Thus the
power-law tails in $P(G_{\Delta t})$ appear to originate from the
power-law tail in $P(w_{\Delta t})$.

Next, consider the long-range correlations found for $\vert G_{\Delta
t} \vert$ \cite{Ding93,Yanhui97,Lundin99,Andersen,Sornette}. Above, we reported
that $N_{\Delta t}$ displays long-range correlations, whereas
$w_{\Delta t}$ does not (Figs.~1--2). Therefore, the long range
correlations in $\vert G_{\Delta t} \vert$ should arise from those
found in $N_{\Delta t}$. Hence, while the power-law tails in
$P(G_{\Delta t})$ are due to the power-law tails in $P(w_{\Delta t})$,
the long-range correlations of $\vert G_{\Delta t} \vert$ are due to
those of $N_{\Delta t}$.

%%%%%%%%%%%%%%%%%%%%%%%%%
In sum, we have shown that stock price movements are analogous to a
complex variant of classic diffusion. Further, we have empirically
demonstrated the relation between stock price changes and market
activity, i.e., the price change at any time $G_{\Delta t}(t)$ is
consistent with a Gaussian-distributed random variable with a local
variance $N_{\Delta t}\,w^2_{\Delta t}$. What could be the
interpretations of our results for the number of transactions
$N_{\Delta t}$ and the local standard deviation $w_{\Delta t}$?  Since
$N_{\Delta t}$ measures the trading activity for a given stock, it is
possible that its power-law distribution and long-range correlations
may be related to
``avalanches''~\cite{Lux99,Cont98,Bouchaud98,Easley92}. The
fluctuations in $w_{\Delta t}$ reflect several factors: (i) the level
of liquidity of the market, (ii) the risk-aversion of the market
participants and (iii) the uncertainty about the fundamental value of
the asset.

%%%%%%%%%%%%%%%%%%%%%%%%%%%%%%%%%%%%%%%%%%%%%%%%%%%%%%%%%%%%%%%%%%%%%%
%%%%%%%%%%%%%%%%%%%%%%%%%%%%%%%%%%%%%%%%%%%% REFERENCES

%%%%%%%%%%%%%%%%%%%%%%%%%%%%%%%%%%%%%%%%%%%%%%%%%%%%%%%%%%%%%%%%%%%%%%
%%%%%%%%%%%%%%%%%%%%%%%%%%%%%%%%%%%%%%%%%%%% FIGURES

{\bf e)} In order to accurately quantify time correlations in
$N_{\Delta t}$, we use the method of detrended
fluctuations~\protect\cite{Peng94} often used to obtain accurate
estimates of power-law correlations. We plot the detrended
fluctuations $F(\tau)$ as a function of the time scale $\tau$ on a
log-log scale for each of the 6 groups. Absence of long-range
correlations would imply $F(\tau) \sim
\tau^{0.5}$, whereas $F(\tau) \sim \tau^{\nu}$ with $0.5 < \nu \leq 1$ 
show power-law correlations with long-range persistence. For each
group, we plot $F(\tau)$ averaged over all stocks in that group. In
order to detect genuine long-range correlations, the U-shaped intraday
pattern for $N_{\Delta t}$ has been removed by dividing each
$N_{\Delta t}$ by the intraday pattern~\protect\cite{Yanhui97}.  For
$0.5 < \nu < 1.0$, correlation function exponent $\nu_{cf}$ and $\nu$
are related through $\nu_{cf} = 2 - 2\nu$. {\bf f)} The histogram of
the exponents $\nu$ obtained by fits to $F(\tau)$ for each of the 1000
stocks shows a relatively narrow spread of $\nu$ around the mean value
$\nu = 0.85 \pm 0.01$.
%\vfill
%\eject
%\vfill
%\eject

\end{multicols}


\begin{thebibliography}{99} 

\bibitem{Bachelier00} 
Bachelier, L., Th\'eorie de la sp\'eculation. {\it Ann. Sci. \'Ecole
Norm. Sup.}  {\bf III-17}, 21--86 (1900).

\bibitem{TAQ} 
{\it The Trades and Quotes Database}, 24 CD-ROMs for 1994-95,
published by the New York Stock Exchange.

\bibitem{Bouchaud}
Bouchaud, J.-P. \& Potters, M., {\it Theory of Financial Risk\/} (Cambridge
University Press, Cambridge 2000)

\bibitem{Farmer}
Farmer, J.~D., Physicists Attempt to Scale the Ivory Towers of
Finance.  {\it Computing in Science \& Engineering} {\bf 26}, November
Issue.
 
\bibitem{Mandelbrot63} 
Mandelbrot, B.~B., The variation of certain speculative prices. {\it
J. Business} {\bf 36}, 394--419 (1963).

\bibitem{Mantegna95} 
Mantegna, R.~N. \& Stanley, H.~E., Scaling behavior in the dynamics of
an economic index. {\it Nature} {\bf 376}, 46--49 (1995).

\bibitem{Ghasghaie96} 
Ghashgaie, S., Breymann, W., Peinke, J., Talkner, P., \& Dodge, Y.,
Turbulent cascades in foreign exchange markets. {\it Nature} {\bf
381}, 767--770 (1996).

\bibitem{Pagan96} 
Pagan, A., The econometrics of financial markets. {\it J. Empirical
Finance} {\bf 3}, 15--102 (1996).

\bibitem{Lux96} 
Lux, T., The stable Paretian hypothesis and the frequency of large
returns: An examination of major German stocks. {\it Applied Financial
Economics} {\bf 6}, 463-75 (1996).

\bibitem{Loretan94} 
Loretan, M. \& Phillips, P.~C.~B., Testing the covariance
stationarity of heavy-tailed time series. {\it J. Empirical Finance}
{\bf 1}, 211--248 (1994)

\bibitem{Muller90} 
Muller, U.~A., Dacorogna, M.~M. \& Pictet, O.~V., Heavy tails in
high-frequency financial data. {\it A Practical Guide to Heavy Tails},
Adler, R.~J., Feldman, R.~E. \& Taqqu, M.~S. (eds.) 283--311
(Birkh\"{a}user Publishers, 1998).

\bibitem{Gopi99} 
Plerou, V., Gopikrishnan, P., Amaral, L.~A.~N., Meyer, M. \& Stanley,
H.~E., Scaling of the distribution of price fluctuations of individual
companies. {\it Phys. Rev. E.} {\bf 60}, 6519-6529 (1999).

\bibitem{Ding93} 
Ding, Z., Granger, C.~W.~J. \& Engle, R.~F., A long memory property of
stock market returns and a new model. {\it J.  Empirical Finance} {\bf
1}, 83--105 (1993).

\bibitem{Yanhui97} 
Liu, Y., Cizeau, P., Meyer, M., Peng, C.-K \& Stanley, H.~E.,
Quantification of correlations in economic time series. {\it Physica
A} {\bf 245}, 437--440 (1997).

\bibitem{Lundin99} 
Lundin, M., Dacorogna, M.~M., \& Muller, U.~A., in {\it Financial
Markets Tick by Tick,} P.~Lequeux (ed.), 91--126, (John Wiley \& Sons,
1999).

\bibitem{Andersen}
Anderson, T., Bollerslev, T., Diebold, F. \& Labys, P., The
distribution of exchange rate volatility. NBER Working Paper WP6961
(1999).

\bibitem{Sornette} 
Arnoedo, A., Muzy, J.-F. \& Sornette, D., Causal cascade in the stock
market from the ``infrared'' to the ``ultraviolet''. {\it
Eur. Phys. J. B} {\bf 2}, 277--282 (1998).

\bibitem{Chandrasekhar}
Chandrasekhar, S., Stochastic problems in Physics and Astronomy. {\it
Rev. Mod. Phys.} {\bf 15} 1--91 (1943), in {\it Selected Papers on
Noise and Stochastic Processes}, Wax, N. (ed.) (Dover Publications
Inc., New York, 1954).

\bibitem{Montroll}
Montroll, E.~W. \& Shlesinger, M.~F., The wonderful world of random
walks, in {\it Nonequilibrium Phenomena II. From Stochastics to
Hydrodynamics} Lebowitz, J.~L. \&  Montroll, E.~W. (eds.) 1-121
(North-Holland, Amsterdam, 1984).

\bibitem{Bouchaud90}
Bouchaud, J.-P. \& Georges, A., Anomalous diffusion in disordered
media: statistical mechanisms, models, and physical applications. {\it
Phys. Rep.} {\bf 195}, 127 (1990).

\bibitem{Clark73} 
Clark, P.~K., A subordinated stochastic process model with finite
variance for speculative prices. {\it Econometrica} {\bf 41}, 135--155
(1973).

\bibitem{Mandelbrot67}
Mandelbrot, B.~B. \& Taylor, H., On the distribution of stock price
differences. {\it Operations Research} {\bf 15}, 1057--1062 (1962).
  
\bibitem{Epps76} 
Epps, T.~W. \& Epps, M.~L., The stochastic dependence of security
price changes and transaction volumes: Implications of the
mixture-of-distributions hypothesis. {\it Econometrica} {\bf 44},
305-321 (1976).

\bibitem{Tauchen83}
Tauchen, G. \& Pitts, M., The price variability-volume relationship on
speculative markets. {\it Econometrica} {\bf 57}, 485--505 (1983).

\bibitem{Stock88}
Stock, J., Estimating continuous time processes subject to time
deformation. {\it Journal of the American Statistical Association}
{\bf 83} 77--85 (1988).

\bibitem{Guillaume}
Guillaume, D.~M., Pictet, O.~V., Muller, U.~A. \& Dacorogna, M.~M.,
Unveiling non-linearities through time scale transformations. Olsen
group preprint OVP.1994-06-26 (1995), available at
http://www.olsen.ch.

\bibitem{Feller} Feller, W., {\it An Introduction to Probability Theory and 
its Applications} (John Wiley, New York, 1966).

\bibitem{Engle95} 
Engle, R.~F., (ed.) {\it ARCH: Selected Readings} (Oxford University
Press, Oxford, 1995).

\bibitem{Ghysels96}
Ghysels, E., Harvey, A. \& Renault, E., Stochastic Volatility. {\it
Handbook of Statistics Vol.~14, Statistical Methods in Finance},
G.~S.~Maddala (ed.), 119--191 (North-Holland, Amsterdam, 1996).


\bibitem{Jones94}
Jones, C., Gautam, K. \& Lipson, M., Transactions, volumes and
volatility.  {\it Reviews of Financial Studies} {\bf 7}, 631--651
(1994).

\bibitem{Ane00}
Ane, T. \& Geman, H., Order flow, transaction clock and normality of
asset returns. {\it Journal of Finance}, forthcoming (2000).

\bibitem{Lux99}
Lux, T. \& Marchesi, M., Scaling and criticality in a stochastic
multi-agent model of a financial market. {\it Nature\/} {\bf 397},
498--500 (1999).

\bibitem{Cont98}
Cont, R. \& Bouchaud, J.-P., Herd behavior and aggregate fluctuations
in financial markets. {\it Macroeconomic Dynamics}, forthcoming;
preprint cond-mat/9712318.

\bibitem{Bouchaud98}
Bouchaud, J.-P. \& Cont, R., A Langevin approach to stock market
fluctuations and crashes. {\it Eur. Phys. J. B} {\bf 6}, 543--550
(1998).

\bibitem{Easley92}
Easley, D. \& O'Hara, M., Time and the process of security price
adjustment. {\it Journal of Finance} {\bf 47}, 905--927.

\bibitem{Peng94} 
Peng, C.-K., {\it et. al.}, Mosaic organization of DNA nucleotides.
{\it Phys. Rev. E} {\bf 49}, 1685--1689 (1994).

 
\end{thebibliography}
\end{document}